\documentclass[prb,onecolumn,preprintnumbers,amsmath,amssymb,11pt]{revtex4}
\usepackage[utf8]{inputenc}
\usepackage{epsf}
\usepackage{graphicx}
\usepackage{latexsym}
\usepackage{amsmath}
\usepackage{amssymb}
\usepackage{amsfonts}
\usepackage{color}

\usepackage{enumitem}

\begin{document}

\title{\Large{Oscillatory Bias Dependence of Visible Height of Monatomic Pb(111) Steps: Consequence of Quantum-Size Effect for Thin Metallic Films}}


\author{Alexey Yu. Aladyshkin$^{(a,b,c,d,*)}$}

\medskip
\affiliation{$^{(a)}$Institute for Physics of Microstructures RAS, GSP-105, 603950 Nizhny Novgorod, Russia \\
$^{(b)}$Lobachevsky State University of Nizhny Novgorod, Gagarin Av. 23, 603022 Nizhny Novgorod, Russia \\
$^{(c)}$National Research University Higher School of Economics (HSE University), Myasnitskaya str. 20, 101000 Moscow, Russia\\
$^{(d)}$Center for Advanced Mesoscience and Nanotechnology, Moscow Institute of Physics and Technology, Institutskiy pereulok 9, 141700 Dolgoprudny, Moscow Region, Russia \\}

\date{\today}

\maketitle




\section{Abstract}

Local structural and electronic properties of thin Pb(111) films grown on Si(111)$7\times 7$ surface are experimentally studied by means of low-temperature scanning tunneling microscopy and spectroscopy (STM/STS). It is shown that the visible height $h$ of the monatomic step on Pb(111) surface demonstrates oscillatory dependence on bias voltage $U$. The period of these oscillations coincides with the period of the oscillations of both local tunneling conductance ($dI/dU$) and the rate of the STM tip displacement ($dZ/dU$) at sweeping $U$. It points to the fact that the observed oscillations of the visible height of monatomic Pb(111) step are controlled by coherent resonant tunneling of electrons from the STM tip to Pb(111) film through quantum-well states in thin Pb(111) film. We argue that the maximum and minimum visible heights of the monatomic Pb(111) step correspond to the bias voltages, at which local densities of states for the Pb(111) terraces of different thicknesses are equal.

$^*$ Corresponding author, e-mail address: aladyshkin@ipmras.ru

\section{Introduction}

An appearance of discrete quantum-well states due to the confinement of electrons in low-dimensional samples is known to determine unusual thermodynamic and transport properties of solid-state nanostructures (see, \emph{e.g.}, textbooks \cite{Ferry-book-09,Nazarov-book-09}). Indeed, for thin-film samples electrons in the conduction band can be considered as nearly free particles for motion in the lateral direction and confined for motion in the transverse direction similar to a 'particle-in-the-box' problem. For such systems, there are localized solutions of the Schr\"odinger equation, which have form of standing electron waves with integer number $n$ of half-waves inside thin conducting film with exponentially decaying tails in the barrier area. Quantization of the transverse wave vector $k^{\,}_{\perp,n}$ and the energy $E^{\,}_{n}$ for the electronic quantum-well states apparently results in the modification of the energy spectrum. In particular, coherent resonant tunneling \cite{Ferry-book-09} through quasi-stationary quantum-well states (see schematic energy diagram in figure~1 from Ref.~3) leads to the appearance of almost equidistantly positioned peaks on the dependence of the differential tunneling conductance $dI/dU$ on the bias voltage $U$ both for filled and empty electronic states in experiments involving scanning tunneling microscopy and spectroscopy (STM/STS). \cite{Altfeder-PRL-97,Altfeder-PRL-98a,Altfeder-PRL-98b,Su-PRL-01,Eom-PRL-06,Hong-PRB-09,Kim-SurfSci-15,Ustavshchikov-JETPLett-17,Putilov-JETPLett-19,Aladyshkin-JPCC-21}

Two-dimensional Pb(111) islands seem to be convenient objects for studying quantum-size effects in metallic films in normal and superconducting states by means of STM/STS \cite{Altfeder-PRL-97,Altfeder-PRL-98a,Altfeder-PRL-98b,Su-PRL-01,Eom-PRL-06,Hong-PRB-09,Kim-SurfSci-15,Ustavshchikov-JETPLett-17,Putilov-JETPLett-19,Aladyshkin-JPCC-21} and by transport and Hall measurements;\cite{Miyata-PRB-08,Jalochowski-PRB-88a,Jalochowski-PRB-88b} for the investigation of electronic properties in superconducting nanostructures \cite{Cren-PRL-09,Ning-ERL-09} and superconducting two-dimensional materials,\cite{Brun-SuST-17} hybrid structures superconductor-ferromagnet,\cite{Aladyshkin-PRB-11a,Aladyshkin-PRB-11b,Iavarone-NatCom-14} superconductor-normal metal,\cite{Cherkez-PRX-14,Roditchev-NatPhys-15} and superconductor-topological insulator;\cite{Stolyarov-JPCLett-21} and for the investigation of electronic states by means of photoemission electron spectroscopy. \cite{Mans-PRB-02,Dil-PRB-06} The particular sensitivity of the interference patterns to the variations of the film thickness and the crystalline structure of the interfaces makes it possible to visualize monatomic steps at the upper/lower interfaces, \cite{Altfeder-PRL-97,Altfeder-PRL-98a,Altfeder-PRL-98b,Kim-SurfSci-15,Ustavshchikov-JETPLett-17} atomic lattice of the substrate covered by metal, \cite{Altfeder-PRL-98a,Altfeder-PRL-98b} various inclusions, \cite{Ustavshchikov-JETPLett-17} terraces with nonquantized height variations, \cite{Putilov-JETPLett-19} and subsurface dislocation loops.\cite{Aladyshkin-JPCC-21}

This paper is devoted to the experimental investigations of structural and electronic properties of thin Pb(111) films in the presence of quantum-well states. One can expect that the periodic variations of the local differential tunneling conductance $dI/dU$ for atomically flat Pb(111) terraces as a function of tunneling voltage $U$ will be converted into an oscillatory dependence of a distance between the STM tip and a sample surface during scanning process. Since the bias-induced variation of the local conductance for the terraces of the thickness differing by one monolayer occurs in out-of-phase manner (see, \emph{e.g.,} Refs. \cite{Ustavshchikov-JETPLett-17,Putilov-JETPLett-19}), one can anticipate to detect periodic variations of the visible height of the monatomic step on the upper surface of the Pb(111) film as sweeping $U$. To the best of our knowledge, we report on a new and reproducible effect concerning a direct experimental observation of the oscillatory dependence of the apparent height of the monatomic step between two atomically flat terraces on top of the Pb(111) island at varying tunneling voltage. We believe that this finding can be of scientific and methodological interest as a clear example of an effect of the quantum-well states on topography images of thin metallic film. We think that the effect described in our paper is rather instrumental caused by the systematic influence of quantum-well states on resulting tunneling conductance than real field-induced expansion deformation in ultrathin Pb islands in a strong electric field.\cite{Chan-PRL-12}

\section{Methods}

Experimental investigations of structural and electronic properties of quasi-two-dimensional Pb islands were carried out in an ultra-high vacuum (UHV) low-temperature scanning probe microscopy setup (Omicron Nanotechnology GmbH) operating at a base vacuum pressure \mbox{$2\cdot10^{-10}\,$mbar}. Si(111) crystals were first out-gassed at about 600$^{\circ}$C for several hours and then cleaned \emph{in-situ} by direct-current annealing at about 1300$^{\circ}$C, resulting in formation of reconstructed Si(111)7$\times$7 surface. Thermal deposition of Pb (Alfa Aesar, purity of 99.99\%) from a Mo crucible was performed {\it in situ} on Si(111)7$\times$7 surface at room temperature by means of an electron-beam evaporator (Focus GmbH, model EFM3) at \mbox{$6\cdot10^{-10}\,$mbar}. The orientation of the atomically flat terraces at the upper surface of the Pb islands corresponds to the (111) plane \cite{Altfeder-PRL-97,Su-PRL-01,Eom-PRL-06}. All STM/STS measurements were carried out at liquid nitrogen temperatures (from 77.4 to 80\,K) with electrochemically etched W tips cleaned {\it in situ} by electron bombardment.

The topography of the Pb islands was studied by low-temperature STM by tracking the displacement of a STM tip mounted on a piezoscanner during scanning above the sample surface with active feedback loop (at constant tunneling current $I$) and constant electrical potential of the sample ($\varphi^{\,}_s=U$) with respect to a virtually grounded STM tip ($\varphi^{\,}_t=0$). Hereafter we will use the following correction coefficients $k^{\,}_{\|}\simeq 1.14$ and $k^{\,}_{\perp}\simeq 1.14$ for scanning tunneling measurements in the lateral and transverse directions, respectively. These factors convert nominal values $x'$, $y'$ and $z'$ values, recorded by a STM Control Unit (raw data), into real dimensions: $x=x'\cdot k^{\,}_{\|}$, $y=y'\cdot k^{\,}_{\|}$ and $z=z'\cdot k^{\,}_{\perp}$. The procedure of fine calibration of piezoscanner is described in Supporting Information in detail. The local electronic properties of the Pb islands were studied by low-temperature STS in the regime with active feedback, constant tunneling current, and variable distance between the tip and the sample surface.\cite{Aladyshkin-JPCC-22}

We would like to emphasize that the signal of the feedback loop acquired in the regime of the constant tunneling current is usually interpreted as a topography map $z = z(x, y)$. We demonstrate below that the signal of the feedback loop is not solely determined by sample topography, and it is influenced by the local electronic properties of a nanostructured sample.

\section{Results and Discussion}

\begin{figure*}[h!]
\centering{\includegraphics[width=16cm]{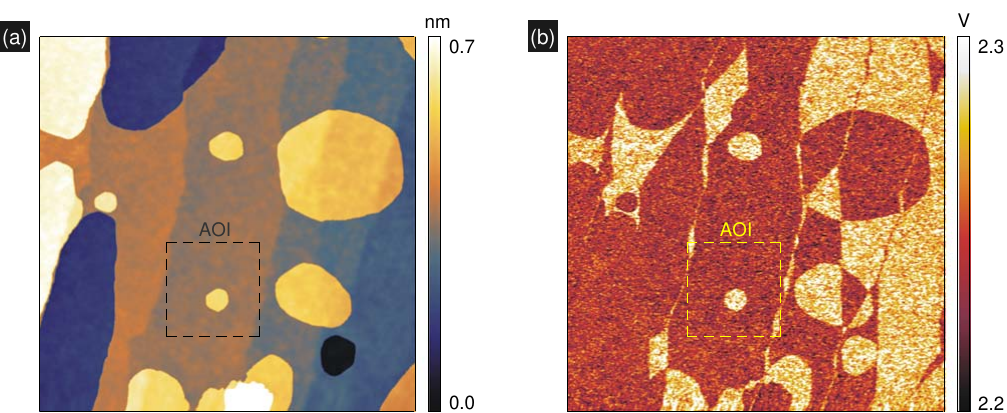}}
\caption{(a) Aligned topographical image of Pb(111)/Si(111)$7\times 7$ film with several atomically flat terraces (image size $456\times 456$\,nm$^2$ after correction, bias voltage $U=0.75$\,V, mean tunneling current $I=400$\,pA, temperature $T=78.5\,$K). (b) Map of local differential tunneling conductance acquired simultaneously with the topography image (bias voltage $U=0.75$\,V, frequency and amplitude of bias modulation are 7285 Hz and 40 mV, respectively). The inner part of the area of interest (AOI, image size $114\times 114$\,nm$^2$ after correction) marked by dashed rectangle does not contain any visible and hidden defects.}
\label{Fig-01}
\end{figure*}

\begin{figure*}[h]
\centering{\includegraphics[width=15cm]{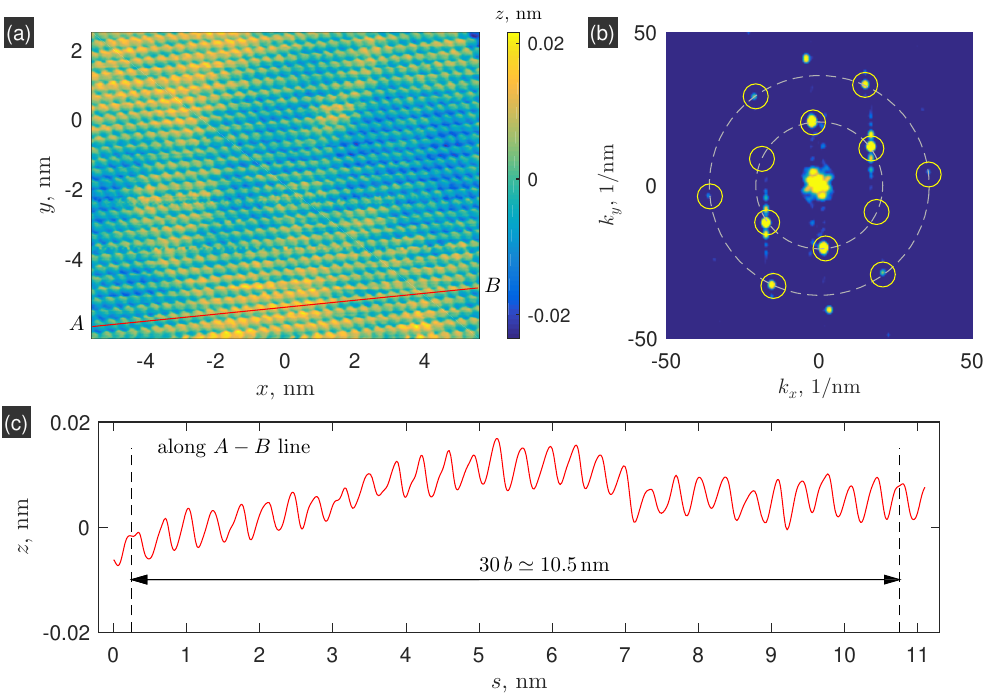}}
\caption{(a) Aligned topography image of clean Pb(111) terrace, showing periodic small-scale corrugation on top of large-scale variations of the local thickness of Pb(111) film (image size 11.1\,$\times 8.8$\,nm$^2$ after correction, bias $U=-0.1$\,V, current $I=40$\,pA, temperature 77.5\,K). (b) Structure of $k-$space obtained by fast Fourier transformation for the image in panel a. This evidences for the formation of a two-dimensional hexagonal lattice, circles mark the Fourier peaks of the first and second orders.  (c) Profile along the $A-B$ line (red line in panel a), the period of oscillations $b\simeq 0.35$\,nm, and typical amplitude of corrugation is about 0.01\,nm.}
\label{Fig-01b}
\end{figure*}

An overview topography image of the Pb(111) film with several atomically flat terraces on the upper surface is presented in figure~\ref{Fig-01}a. Hereafter, we focus on local structural and electronic properties of the area (depicted by a dashed rectangle) containing two flat terraces, whose local thicknesses differ by one monolayer. Comparing the topography image $z(x,y)$ (figure~\ref{Fig-01}a) and the map of local differential conductance $dI/dU(x,y)$ (figure~\ref{Fig-01}b) recorded simultaneously at the same bias voltage, we conclude that there are no visible and hidden defects like monatomic steps in Si(111) substrate and/or inclusions inside the area of interest (AOI). It ensures that the compensation of the global tilt of the sample on the topography images both along $x-$ and $y-$axes can be performed with high accuracy. As a consequence, the estimate of the visible height of monatomic Pb(111) step within such AOI seems to be the most reliable. A typical small-scale topography image demonstrates a two-dimensional hexagonal lattice for a single terrace of Pb(111) film (figure~\ref{Fig-01b}a).

\begin{figure*}[ht!]
\centering{\includegraphics[width=16cm]{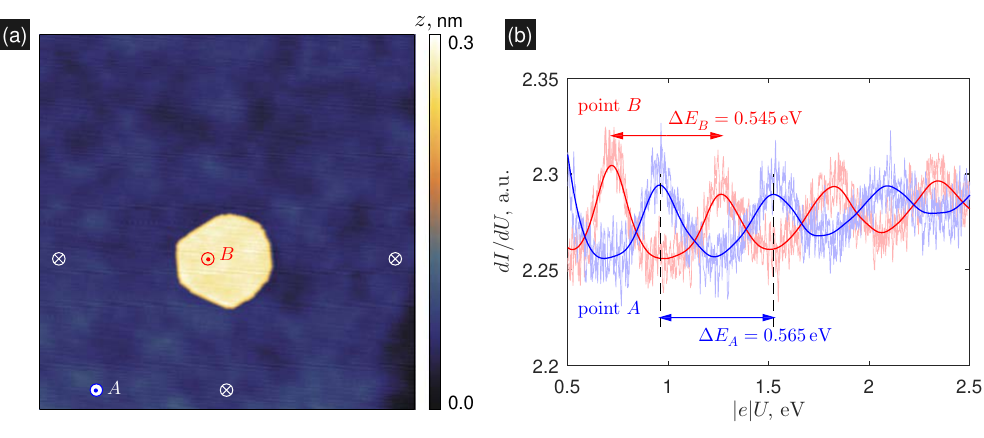}}
\caption{(a) Aligned topography image within the area of interest (image size $114\times 114$\,nm$^2$ after correction, $U=0.50$\,V, $I=400$\,pA, temperature 79.5\,K). Removal of the global background tilt of the sample was done by the subtraction of a plane defined by three reference points $\otimes$. (b) Bias dependence of the local differential tunneling conductance $dI/dU$ for two points $A$ and $B$ (see panel a). The periods of the quantum-size oscillations $\Delta E$ are equal to $0.565\,$eV (point $A$, outside the circular terrace) and $0.545\,$eV (point $B$, inside the circular terrace). It gives us the estimates for the local thickness of the Pb(111) film: 23 monolayers for the lower (blue) terrace and 24 monolayers for the upper (yellow) terrace.}
\label{Fig-02}
\end{figure*}

In order to investigate experimentally the dependence of the visible height of monatomic Pb(111) step on tunneling voltage $U$ we carry a series of 24 combined topographic-spectroscopical measurements within the area of interest. Each measurement takes about 20 min and gives us the topography map $z(x,y)$ and the map of the differential tunneling conductance $dI/dU(x,y)$, both acquired at the same $U$ value. Since clean surfaces of the Pb(111) film are not ideally flat and large-scale inhomogeneities are detectable (see yellow and blue regions in figure~\ref{Fig-01b}a and blue and dark-blue regions in figure~\ref{Fig-02}a), there is a question concerning a reliable estimate of the height of the monatomic Pb(111) step. In addition, due to unavoidable thermal drift, the particular location of the upper terrace with respect to the bounds of the scanning area changes during long-term investigations (24 'forward-and-backward' scans within one working day). The drift makes impossible simple mathematical subtraction one topography image recorded at bias voltage $U^{\,}_1$ from another image recorded at bias voltage $U^{\,}_2$ to find any differences, related to the effect of bias voltage. To avoid additional instrumental complications, we consider only forward measurements.

Figure~\ref{Fig-02}a shows typical topography image within the area of interest. The dependences of the local differential tunneling conductance $dI/dU$ on $U$ acquired for the terraces of different thicknesses (points $A$ and $B$) in the regime with active feedback loop are presented in figure~\ref{Fig-02}b. Pronounced periodic oscillations in the dependence $dI/dU$ on $U$ at low bias voltage ($U \lesssim 3$V) can be considered as clear experimental evidence for the coherent resonant tunneling through quantum-well states localized in thin metallic film \cite{Altfeder-PRL-97,Altfeder-PRL-98a,Altfeder-PRL-98b,Su-PRL-01,Eom-PRL-06,Hong-PRB-09,Kim-SurfSci-15,Ustavshchikov-JETPLett-17,Putilov-JETPLett-19,Aladyshkin-JPCC-21}. Taking the periods of the quantum-size oscillations $\Delta E^{\,}_A\simeq 0.565\,$eV and $\Delta E^{\,}_B\simeq 0.545\,$eV, one can estimate the local thickness of the Pb film with respect to the Si(111)$7\times 7$ surface using the relationships $D^{\,}_A \simeq \pi\hbar v^{\,}_F /\Delta E^{\,}_A\simeq 6.57\,$nm and $D^{\,}_B \simeq \pi\hbar v^{\,}_F /\Delta E^{\,}_B\simeq 6.86\,$nm, where $v^{\,}_F=1.8\cdot 10^8$\,cm/s is the Fermi velocity for the Pb(111) films. The theoretical estimate for the monatomic step on the Pb(111) surface is $d^{\,}_{ML}=a/\sqrt{3}=0.286\,$nm, where $a=0.495\,$nm is the lattice constant for bulk Pb. As a consequence, the local thickness for the lower terrace in figure~\ref{Fig-02}a is equal to 23 monolayers, while the local thickness for the upper terrace is equal to 24 monolayers.

\begin{figure*}[t!]
\centering{\includegraphics[width=7.5cm]{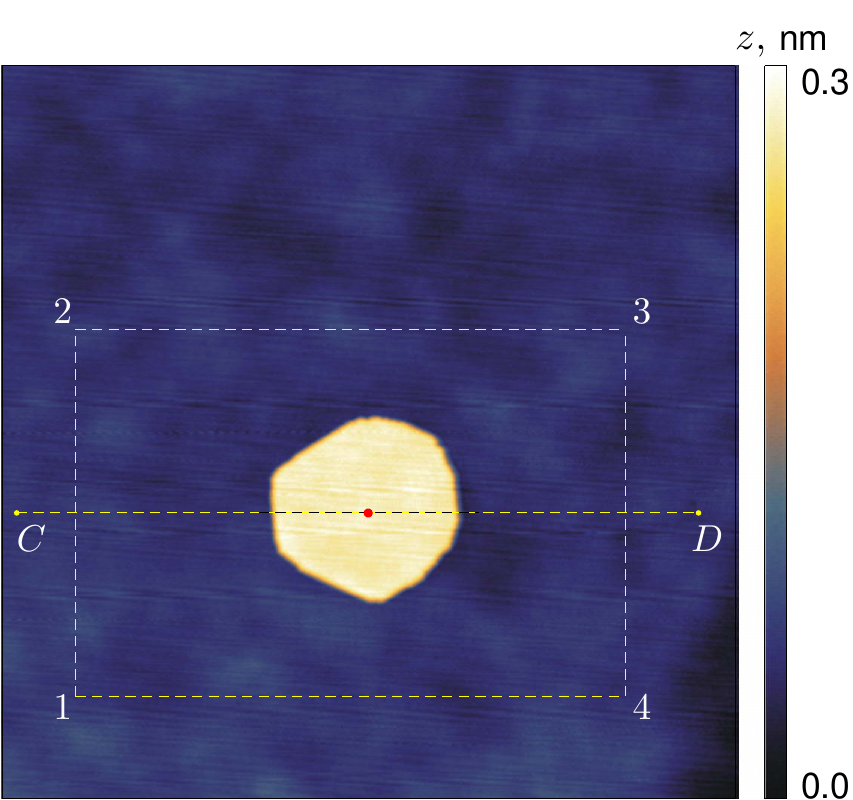}}
\caption{Same topography image as in figure \ref{Fig-02}a. Contour 1-2-3-4 around the upper terrace is used for selection the data set for further statistical analysis and plotting histograms (figure~\ref{Fig-04}a). Line $C-D$ running through the center of the upper terrace (red point) is used for plotting cross-sectional views (figure~\ref{Fig-05}a). }
\label{Fig-03}
\end{figure*}

In order to estimate the apparent height of the monatomic Pb(111) step we apply three different approaches.

\begin{figure*}[ht!]
\centering{\includegraphics[width=16cm]{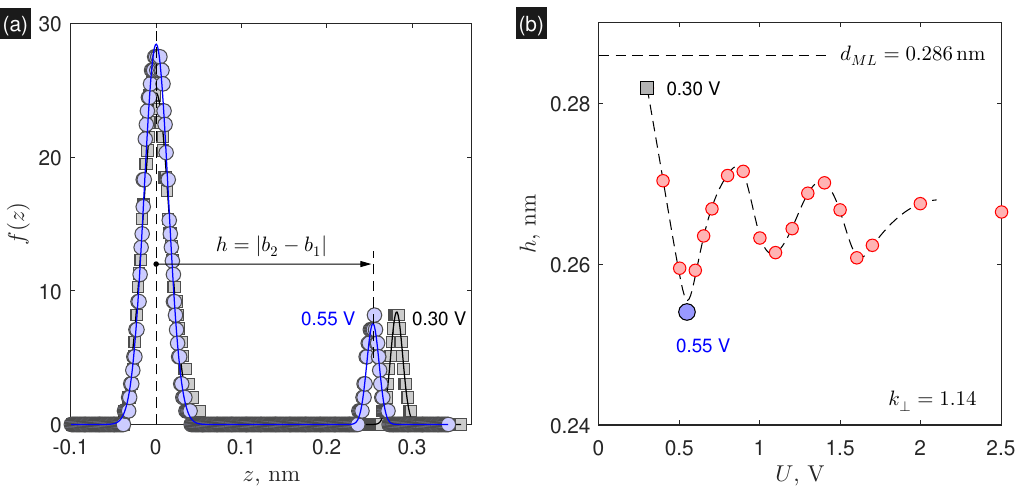}}
\caption{(a) Normalized probability density functions $f(z)$, illustrating the height distribution within the contour 1-2-3-4 (see figure \ref{Fig-03}) for $U=0.300\,$V (gray squares, maximum visible height) and $U=0.550\,$V (blue circles, minimum visible height). (b) Dependence of the visible height $h$ of the monatomic Pb(111) step, defined as the interval between two maxima in the $f(z)$ distribution, on bias voltage $U$. During this series of a 16-hour measurement, a temperature of the sample increases at 0.6\,K (from 79.4 to 80.0\,K). The size of the symbols is close to the confidence interval for the $h$ values. The horizontal dashed line marks the theoretical limit for the height of the Pb(111) monolayer. }
\label{Fig-04}
\end{figure*}

\medskip
First, we manually choose a rectangular area of the same size surrounding the upper terrace for each topography image (see contour 1-2-3-4 in figure~\ref{Fig-03}). Then, we compose histograms illustrating the height distribution within this area for each scan. It is obvious that the probability density function $f(z)$ should have two maxima (figure~\ref{Fig-04}a): the positions of the main and minor peaks correspond to the mean relative heights for the lower and upper terraces, respectively. The probability density function $f(z)$ for each $U$ value can be well fitted by a superposition of two Gaussian functions
\begin{eqnarray}
\label{Eq-Two-Gaussians}
f^{\,}(z) = a^{\,}_1\cdot \exp\left(-\frac{(z-b^{\,}_1)^2}{2c^{2}_1}\right) + a^{\,}_2\cdot \exp\left(- \frac{(z-b^{\,}_2)^2}{2c^{2}_2}\right),
\end{eqnarray}
where $a^{\,}_{1,2}$, $b^{\,}_{1,2}$ and $c^{\,}_{1,2}$ are fitting parameters. Measuring the interval between these peaks for the series of measurements at different bias voltages, one can determine the dependence of the visible height of the monatomic Pb(111) step: $h\equiv |b^{\,}_2-b^{\,}_1|$ on $U$ (figure~\ref{Fig-04}b).

Considering data presented in figure~\ref{Fig-04}, we conclude that

(i) the visible height of the monatomic Pb(111) step varies with oscillations as bias voltage monotonously increases;

(ii) the maximal effect of the bias-induced variation of the height observed experimentally is about 0.03 nm or 10\% from the theoretical limit for the Pb(111) monolayer (0.286 nm);

(iii) the detected period of the height oscillations (about 0.55 eV) is very close to the period of the variations of the differential tunneling conductance $dI/dU$ as a function of $U$, controlled by quantum-well states in thin Pb films (figure~\ref{Fig-02}a);

(iv) all $h$ values are smaller than the theoretical limit for the Pb(111) monolayer, provided we use the mean correction factor $k^{\,}_{\perp}=1.14$.

\begin{figure*}[t!]
\centering{\includegraphics[width=16cm]{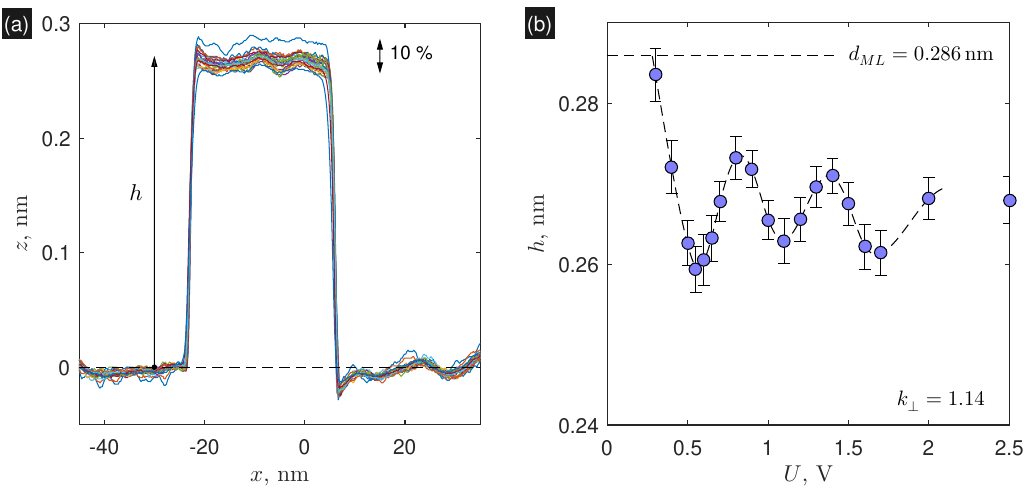}}
\caption{(a) Set of the profiles taken along the $C-D$ line running through the center of the upper terrace for dufferent bias voltages. (b) Oscillatory dependence of $h$ on $U$ for the same set of raw data as in figure~\ref{Fig-04}b. The errorbars are proportional to the standard deviation of real profiles from the corresponding mean values and meet to 95\% confidence level.}
\label{Fig-05}
\end{figure*}

\medskip
In order to confirm that the observed oscillatory dependence of $h$ on $U$ is not an artifact of statistical treatment, we consider alternative methods of data analysis.

\medskip
The second approach is based on a cross-sectional analysis. We find the geometrical center of the upper terrace (red point in figure~\ref{Fig-03}) and then plot the cross-sectional view along the $x-$axis via the center of the terrace regardless of particular locations of this terrace with respect to bounds of a topography image for all scans. All these profiles are shown in figure~\ref{Fig-05}a. The visible height of the monatomic step $h$ can be defined now as the vertical separation between the mean $z$ values for the upper terrace and for the lower terrace. It is easy to see that the mean height of the monatomic step with respect to the background level indeed depends on the bias voltage. The dependence of $h$ on $U$ shown in figure~\ref{Fig-05}b expectedly coincides with the result of the histogram analysis (figure~\ref{Fig-04}b).

\begin{figure*}[t!]
\centering{\includegraphics[width=16cm]{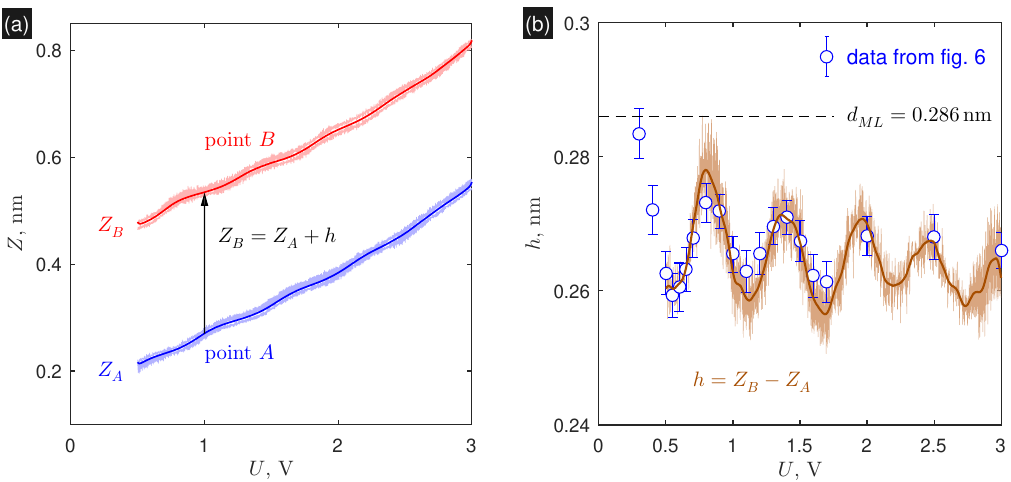}}
\caption{(a) Two local $Z-U$ spectra, recorded at points $A$ and $B$ (the lower and upper terraces, correspondingly, see figure~\ref{Fig-02}), tunneling current $I=400\,$pA, temperature 78.9\,K. Thick red and blue lines are the mean $Z-U$ spectra averaged over a series of three measurements and then smoothed by a running Gaussian filter with a window of 20 mV in order to remove high-frequency noise. (b) Dependence of the visible height of the monatomic step defined as a point-by-point difference of $Z^{\,}_B$ and $Z^{\,}_A$ for raw data for the last measurements in this series as a function of bias voltage $U$ (thin brown line). The thick brown line corresponds to the dependence of $Z^{\,}_B-Z^{\,}_A$ on $U$ averaged over a window of 20 mV. White circles are data of cross-sectional analysis from figure~\ref{Fig-05}b.}
\label{Fig-06}
\end{figure*}

\medskip
Third, we acquire a series of the local distance-voltage ($Z-U$) spectroscopic measurements in the regime of constant tunneling current.\cite{Aladyshkin-JPCC-22} Two tunneling spectra $Z^{\,}_A-U$ and $Z^{\,}_B-U$ for the lower and upper terraces are shown in figure~\ref{Fig-06}a. It is clear that the dependence of $h=Z^{\,}_B-Z^{\,}_A$ on $U$ (figure~\ref{Fig-06}b) very well reproduces the results of previous analysis (see figures~\ref{Fig-04} and~\ref{Fig-05}). It is important to note that $Z-U$ measurements are rather fast (1 min per line for entire range of the considered $U$ values instead of 20 min per two-dimensional scan for a single $U$ value), therefore temperature-induced variations of both a piezocoefficient and the correction factor $k^{\,}_{\perp}$ should be negligible (see figures S2-S4 in the Supporting Information).

\begin{figure}[t!]
\centering{\includegraphics[width=14.5cm]{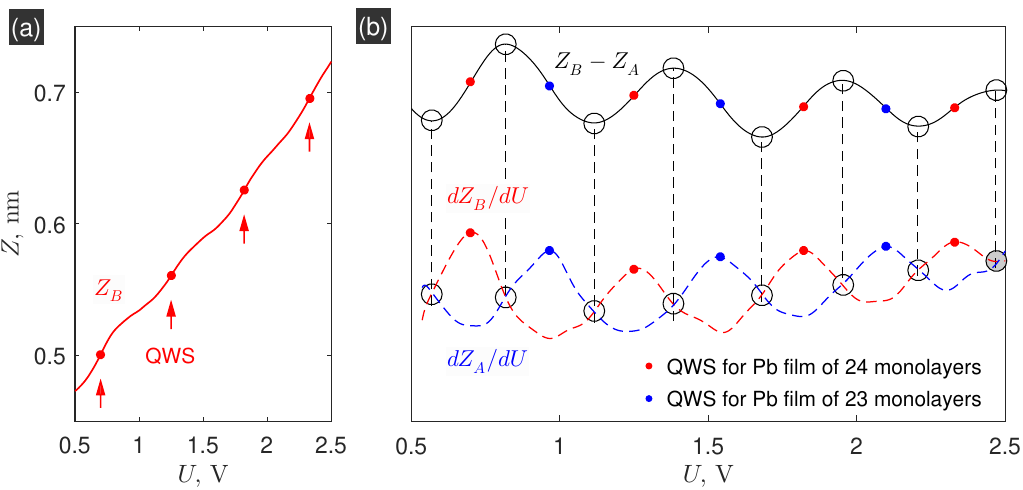}}
\caption{(a) $Z-U$ spectrum for point $B$ (the same dependence is shown in figure~\ref{Fig-06}a as a thick red line); arrows show the $dZ/dU$ maxima, corresponding to the energies of the quantum-well states for this terrace.
(b) Black solid line shows the dependence of $h=Z^{\,}_B-Z^{\,}_A$, averaged over a 40 mV window, as a function of $U$.  Blue and red dashed lines show the bias-dependent variations of the slope $dZ^{\,}_A/dU$ and $dZ^{\,}_B/dU$; both dependences $Z^{\,}_A(U)$ and $Z^{\,}_B(U)$ were averaged over a 40 mV window before numerical differentiation. Red and blue filled dots mark the voltage values, corresponding to the quantum-well states for the lower and upper terraces (\emph{i.e.} to the local maxima in the $dZ^{\,}_A/dU$ and $dZ^{\,}_B/dU$ dependences). Open circles mark the voltage values, corresponding to the extremal values in the dependence of $h$ on $U$.}
\label{Fig-07}
\end{figure}

\medskip
Thus, based on our experimental observations, we conclude that the quantum-well states in thin Pb(111) film play a role of a systematic and reproducible inaccuracy factor affecting the STM measurements in a topographic mode. We would like to emphasize that the actual thickness of the Pb film in the area of interest is not extremely small (23-24 monolayers) and all of the measurements were performed in the limit of low-tunneling current (400\,pA). This substantially differs from Ref.~\cite{Chan-PRL-12}, where properties of ultrathin Pb films (from 2 to 16 monolayers) were studied in the limit of high-tunneling current (up to 50 nA) and high-tunneling voltage (up to 5 V), what a results in strong electric field near the tip apex. As a consequence, we do not consider our bias-dependent oscillatory dependence of the visible height of the Pb (111) step as an indication of the real field-induced deformations of Pb films according to Ref.~\cite{Chan-PRL-12}.

\begin{figure*}[t!]
\centering{\includegraphics[width=15cm]{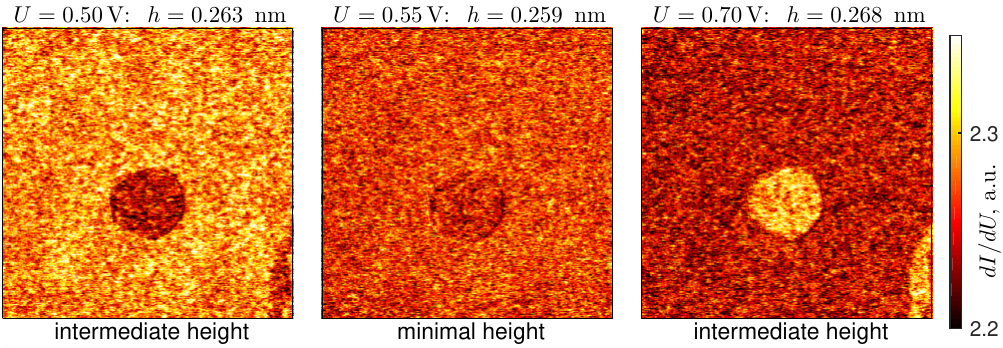}}
\caption{Maps of local differential tunneling conductance $dI/dU(x,y)$ within the area of interest, acquired at bias voltages $U=0.50$, 0.55, and 0.70 V (from left to right) at a temperature range from 79.45 to 79.55\,K (image size $114\times 114$\,nm$^2$ after correction, $I=400\,$pA). Note that we analyze the same set of combined STM/STS measurements, which was already used for preparing figures~\ref{Fig-04}b and \ref{Fig-05}b. The middle image corresponds to the experimental condition of the minimal visible height of the Pb(111) step (compare this value $U=0.55$\,V with the positions of the first minimum in the dependences of $h$ on $U$ presented in figures  \ref{Fig-04}b, \ref{Fig-05}b and \ref{Fig-06}b).}
\label{Fig-08}
\end{figure*}

\medskip
Finally, we would like to discuss the phase of the oscillations of the visible height of the Pb(111) step as a function of $U$ with respect to the $dI/dU-$ and $dZ/dU-$oscillations occurring in phase\cite{Aladyshkin-JPCM-20} and both controlled by resonant tunneling of electrons through quantum-well states in thin Pb(111) films. The black solid line in figure~\ref{Fig-07} shows the bias-induced variations on the visible height $h=Z^{\,}_B - Z^{\,}_A$ derived from experimental $Z^{\,}_A-U$ and $Z^{\,}_B-U$ dependences (figure~\ref{Fig-06}) after the removal of high-frequency noise by additional running Gaussian averaging. Filtration of the noisy component was apparently necessary for the numerical differentiation of raw data and getting the dependences $dZ^{\,}_A/dU$ on $U$ and
$dZ^{\,}_B/dU$ on $U$ (blue and red dashed lines in figure~\ref{Fig-07}b, correspondingly). It is clear that the extremal values of $h$ should correspond to the $U$ values satisfying the conditions
\begin{eqnarray}
\label{Eq:oscillations}
\frac{dh}{dU} = 0 \qquad \mbox{or} \qquad \frac{d Z^{\,}_A}{dU}=\frac{d Z^{\,}_B}{dU}.
\end{eqnarray}
It means that both maximal and minimal values of the $h-$oscillations can be observed at bias voltages, at which the rate of the tip displacement ($dZ/dU$) for the Pb(111) terraces of different thickness are equal (figure~\ref{Fig-07}). Since the oscillations of $dZ/dU$ and $dI/dU$ as a function of $U$ occur in phase, we conclude that the maximal and minimal of the visible height  of the monatomic Pb(111) step should correspond to the case of equal differential tunneling conductance for the Pb(111) terraces of different thicknesses. This conclusion is nicely supported by the results of the combined STM/STS measurements (figure~\ref{Fig-08}).

\section{Conclusion}

We experimentally studied the peculiarities of resonant tunneling through quantum-well states in thin Pb(111) films by means of low-temperature scanning tunneling microscopy and spectroscopy. All measurements were carried out in the regime constant tunneling current with an active feedback loop. We argued that the quantum-well states in Pb(111) films are responsible for the oscillatory dependence of the visible height of the monatomic Pb(111) step on the sample surface, measured by scanning tunneling microscopy, as a function of bias voltage. We experimentally demonstrated that the maximum and minimum visible heights of the monatomic Pb(111) step correspond to the bias voltages, at which there is no contrast in the differential tunneling conductance for the Pb(111) terraces of different thicknesses. We believe that such the oscillatory behavior should be a common property for all thin-film samples, provided that resonant electron tunneling via quantum-well states is not completely suppressed by thermal effects and structural imperfections.

\section*{Supporting Information}
Additional figures illustrating the details of fine calibration of piezoscanner at liquid nitrogen temperatures

\section*{Acknowlednements}

The author is  grateful to S. I. Bozhko, S. V. Zaitsev-Zotov, A. A. Zhukov and A. V. Putilov for fruitful discussions and valuable comments. The work was performed with the use of the facilities at the Common Research Center 'Physics and Technology of Micro- and Nanostructures' at Institute for Physics of Microstructures RAS and funded by the Russian State Contract (No.~FFUF-2021-0020).

\newpage

\setcounter{page}{1}
\setcounter{figure}{0}

\renewcommand{\thefigure}{S\arabic{figure}}
\renewcommand{\thepage}{S\arabic{page}}
\renewcommand{\theequation}{S\arabic{equation}}

\begin{center}
{\large {\bf \textcolor[rgb]{0.00,0.00,0.55}{Supporting Information}}}
\end{center}

\vspace*{0.2cm}

\begin{center}
{\large {\bf Oscillatory Bias Dependence of the Visible Height of Monatomic Pb(111) Steps: Consequence of the Quantum-Size Effect for Thin Metallic Films}}
\end{center}

\vspace*{0.1cm}


%
%
%

\vspace*{1cm}

Reconstruction Si(111)$7\times 7$ (see [S.1]-[S.3]) with well-known parameters can be considered as a test surface for the calibration of our piezoscanner at liquid nitrogen temperatures. We would like to emphasize that for the calibration purposes we use 0.5-mm thick Si(111) single crystals with clean surface without Pb layer on top. All measurements (both main and supporting) were carried out with a thin-wall piezoscanner extended at about 80\,\% from the maximal value after thermal stabilization (at least 3-4 hours before starting measurements).

\section*{Fine calibration of piezoscanner in the vertical direction}

\begin{figure*}[h!]
\centering{\includegraphics[width=13cm]{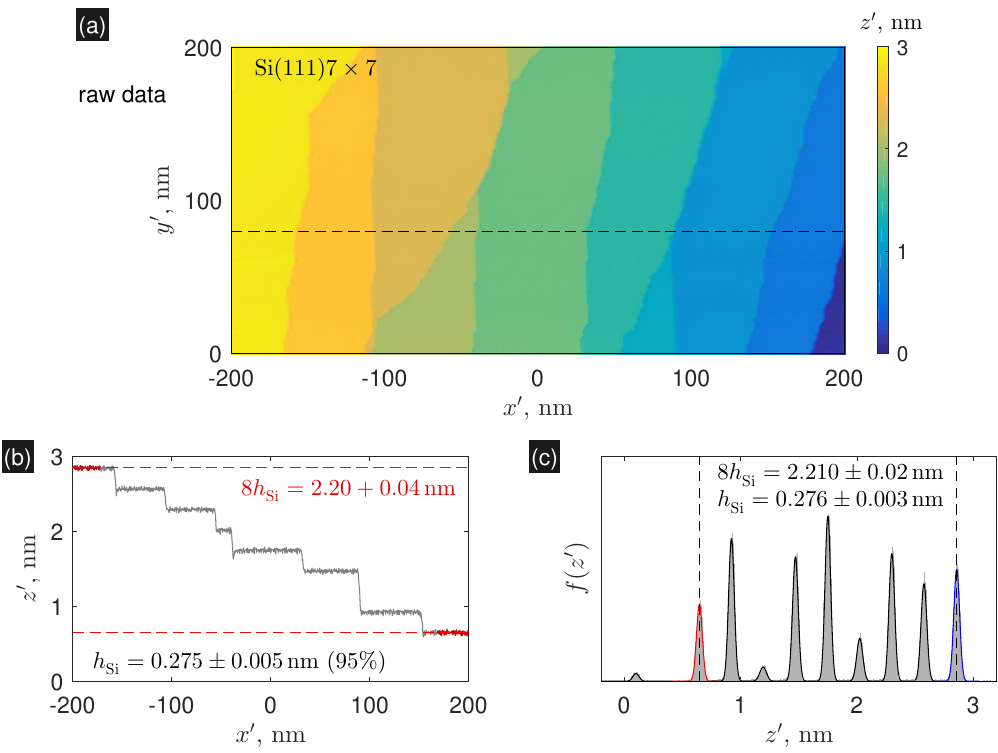}}
\caption{(a) Aligned topography image of annealed Si(111) single crystal with several atomically flat terraces (nominal image size $400\times 200$\,nm$^2$, tunneling voltage $U=0.60$\,V, current $I=100$\,pA, temperature $T=78$\,K). (b) The profile of the surface along the dashed line shown in the panel~a. (c) Probability density function $f(z')$, illustrating the distribution of the nominal heights for the topography image presented in the panel~a. \\ All $x'$ and $z'$ values are non-corrected dimensions (in the other words, output values recorded directly from the STM Control Unit). }
\label{Fig-Si-hist}
\end{figure*}

Figure~\ref{Fig-Si-hist}(a) shows typical topography image recorded at $U=0.6\,$V and $T \simeq 78\,$K. This image was aligned in such a way to remove global tilt both along $x-$ and $y-$axes (\emph{i.e.} along fast and slow scanning directions, respectively). Considering typical cross-sectional view [figure~\ref{Fig-Si-hist}(b)], we estimate the visible height of the monatomic steps on the Si(111)$7\times 7$ surface: $h^{\,}_{\rm Si} = 0.275\pm 0.005$\,nm (with 95\% confidence level). Comparing this value with the theoretical prediction for an ideal Si(111) monolayer $d^{\,*}_{ML}= a/\sqrt{3} =0.3135$\,nm (where $a=0.5431\,$nm is the lattice constant for bulk Si), we conclude that the dimensionless correction factor $k^{\,}_{\perp}$ is equal to
\begin{equation}
k^{\,}_{\perp} = \frac{d^{\,*}_{ML}}{h^{\,}_{\rm Si}} =\frac{0.3135}{0.275\pm 0.005} \simeq 1.140 \pm 0.018.
\label{Eq:k-perp}
\end{equation}
Thus, nominal output values $z'$ can be converted into real dimensions $z$ by the following rule
\begin{equation}
z = z'\cdot k^{\,}_{\perp}.
\label{Eq:k-perp-2}
\end{equation}

\bigskip
If necessary, the $k^{\,}_{\perp}$ value can be refined using statistical analysis applied to two-dimensional images. Figure~\ref{Fig-Si-hist}(c) shows the histogram, which illustrates the distribution of the visible heights for the aligned topography image in figure~\ref{Fig-Si-hist}(a). The composed histogram has ten narrow peaks, corresponding to the contribution of ten atomically-flat terraces. Each of the peaks in the probability density function $f(z')$ can be described by a Gaussian function
\begin{equation}
f^{\,}_n(z') = a^{\,}_n\cdot \exp\left(-\frac{(z'-z^{\,}_n)^2}{2c^2} \right),
\label{Eq:gaussian}
\end{equation}
where $a^{\,}_n$ is the coefficient proportional to the area of the $n-$th terrace,  $z^{\,}_n$ is the mean visible height of the $n-$th terrace and $c$ is the coefficient (independent on $n$), which characterizes natural corrugation for the atomically-flat terrace and the accuracy of the tilt compensation. Comparing the positions of two peaks (marked by red and blue solid lines), one can independently estimate the visible height of the monatomic Si(111) step: $h^{\,}_{\rm Si} = 0.276\pm 0.003$\,nm (with 95\% confidence level). This two-dimensional statistical analysis gives us the better estimate for the correction factor
\begin{equation}
k^{\,}_{\perp} =  \frac{0.3135}{0.276\pm 0.003} \simeq 1.136 \pm 0.015.
\label{Eq:k-perp-2}
\end{equation}

\begin{figure*}[h!]
\centering{\includegraphics[width=11cm]{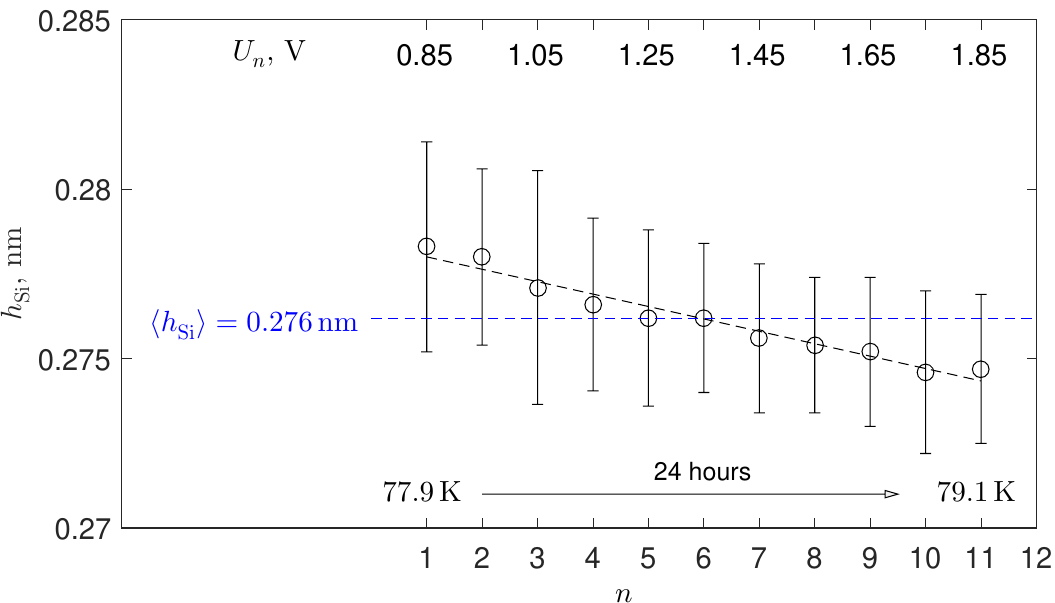}}
\caption{Dependence of the visible height of monatomic Si(111) step on sequential number of measurement $n$. The upper horizontal axis shows the bias voltage $U^{\,}_n$, corresponding to the $n-$th measurement. All measurements in this series were done for the scanning area $400\times 400\,$nm$^2$ before correction, scanning velocity $100\,$nm/s, 70 min per scan. An almost linear dependence of $h^{\,}_{\rm Si}$ on $U$ seems to be illusory correlation, since the dominant effect is a temperature rise at about 1.2\,K occurring during this series of 24-hour measurements (see also figure~\ref{Fig-temp-dependence-2}).}
\label{Fig-temp-dependence}
\end{figure*}

\begin{figure*}[h!]
\centering{\includegraphics[width=11cm]{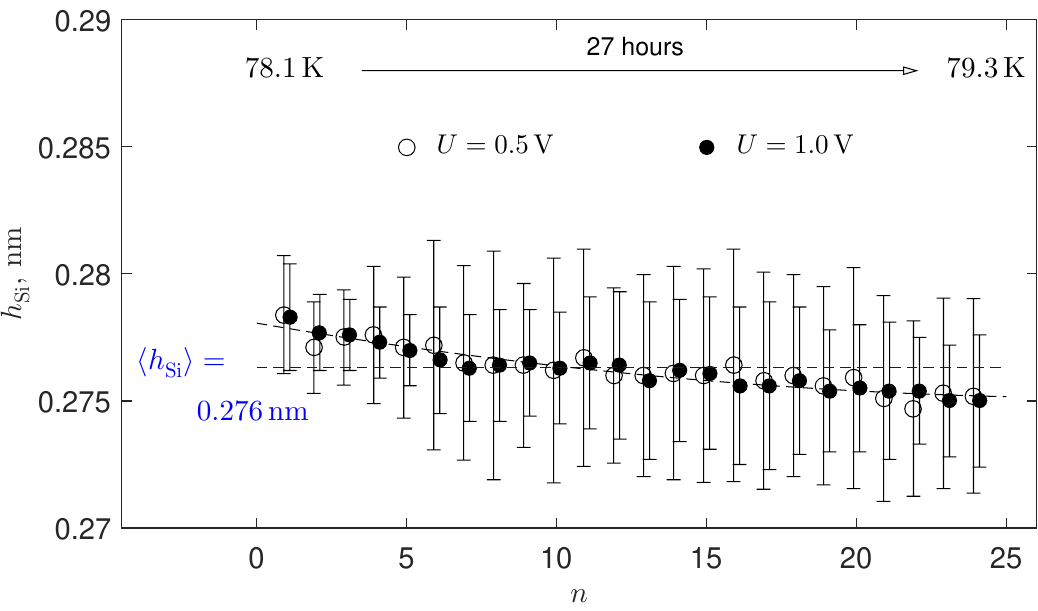}}
\caption{Dependence of the visible height of monatomic Si(111) step on sequential number of measurement $n$. Evolution of the sample temperature, measured by a built-in thermo sensor, is indicated in the plot. All measurements in this series were done for $U=0.5\,$V (forward scanning direction, open circles) and $U=1.0\,$V (backward scanning direction, filled circles), the scanning area $400\times 400\,$nm$^2$ before correction, scanning velocity $100\,$nm/s, 70 min per scan. }
\label{Fig-temp-dependence-2}
\end{figure*}

\bigskip
In order to check whether variations of tunneling voltage and/or unavoidable heating of both sample/piezoscanner during long-term measurements critically affect the correction factor $k^{\,}_{\perp}$, we perform independent series of the topographic measurements for the fixed area of clean Si(111)$7\times 7$ surface. The results of the estimates of the uncorrected visible height of the monatomic Si(111) step by means of the histogram analysis (similar to that shown in figure~\ref{Fig-Si-hist}(c)) are presented in figure~\ref{Fig-temp-dependence}. One can see that the mean visible height $h^{\,}_{\rm Si}$ \emph{monotonously} decreases at 1.3\% during 27-hour measurements (from 0.278 to 0.274 nm). The absence of bias-induced oscillations of $h^{\,}_{\rm Si}$ for bulk sample points to that the oscillatory dependence for the visible height of the Pb(111) step described in the main paper is inherent only to thin-film samples. We think that an almost linearly decreasing dependence of $h^{\,}_{\rm Si}$ on tunneling voltage $U$ should be considered as a false correlation, since the bias voltage is applied directly to the STM tip and therefore it cannot affect the properties of the piezoscanner provided that parasitic cross-talk between measurement sample-tip circuit and feedback loop is absent.

\begin{figure*}[t]
\centering{\includegraphics[width=12.5cm]{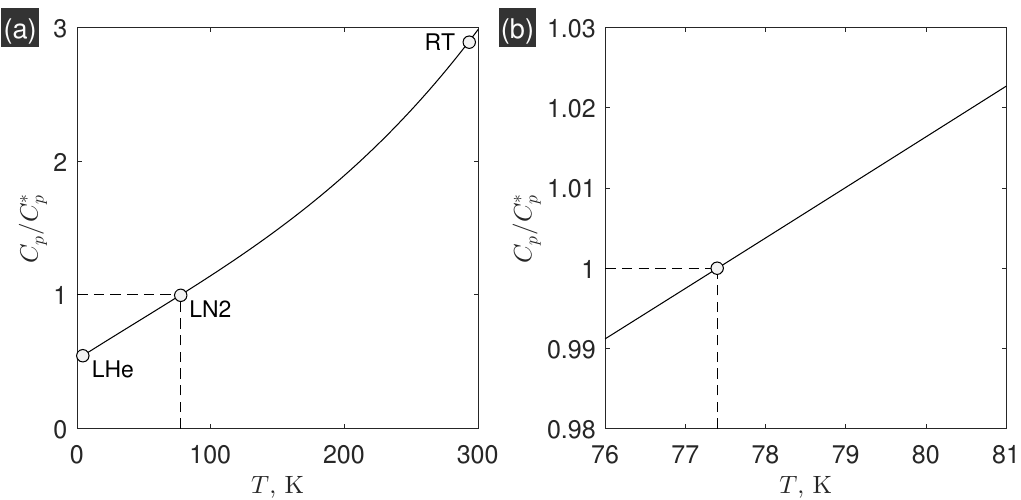}}
\caption{Temperature dependence of piezoelectrical coefficient $C^{\,}_p$, normalized to its value $C^{*}_p$ at $T=77.4\,$K, for PZT ceramics at broad temperature range (a) and at liquid nitrogen temperatures (b), according to data in [S.4]-[S.5].}
\label{Fig-temp-dependence-piezo}
\end{figure*}

To confirm that the observed monotonously decreasing dependence of $h^{\,}_{\rm Si}$ on $n$ unambiguously relates to a temperature rise,  we measure a temporal dependence of $h^{\,}_{\rm Si}$ for the given bias values (figure~\ref{Fig-temp-dependence-2}). During such 27-hour measurements the temperature of the sample was increased at 1.2\,K (from 78.1 to 79.3\,K), resulting in a decrease of $h^{\,}_{\rm Si}$ at 1.5\% (from 0.279 to 0.275 nm) regardless on the tunneling voltage. The described tendency (figures \ref{Fig-temp-dependence} and \ref{Fig-temp-dependence-2}) is in qualitative and quantitative agreement with typical temperature dependence of piezo coefficient (figure~\ref{Fig-temp-dependence-piezo}). Indeed, the increase in temperature of the piezoscanner at 1\,K leads to the increase in the piezo coefficient at about 0.7\%  (see [S.4]-[S.5]); therefore, the feedback loop should generate high voltage, applied to the piezoscanner, at 0.7\% smaller in order to get the same displacement of the STM tip. It is worth noting that the temperature of the piezoscanner may differ from the recorded temperature assigned to the sample.

The monotonous dependence of $h^{\,}_{\rm Si}$ on the sequence number $n$ of measurement can be transformed into a monotonous dependence of the correction factor $k^{\,}_{\perp}$ on temperature. For sake of simplicity we estimate the \emph{mean} correction factor taking into account temperature variations
\begin{equation}
k^{\,}_{\perp} \simeq 1.14 \pm 0.02
\label{Eq:k-perp-3}
\end{equation}
for entire range of tunneling voltages and temperatures considered in the main paper (see figures~5 and 6 of the main paper). The monotonous dependence of $k^{\,}_{\perp}$ on $T$ (figure~\ref{Fig-temp-dependence-2}) confirms that the detected oscillatory dependence of the visible height of the monatomic Pb(111) step on the bias voltage cannot be an artifact of the measurement systems.

\section*{Fine calibration of piezoscanner in the lateral direction}

\begin{figure*}[h]
\centering{\includegraphics[width=12.5cm]{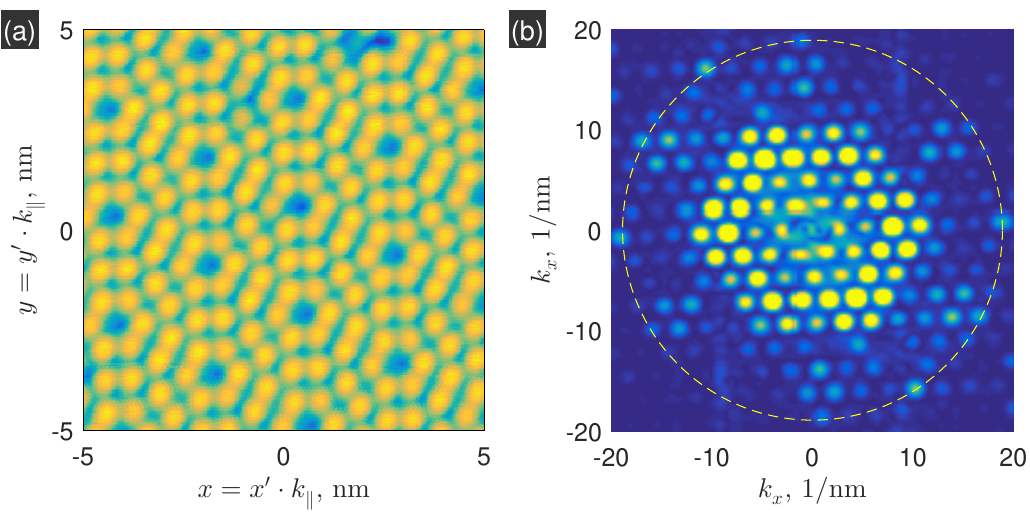}}
\caption{(a) Aligned topography image showing $7\times 7$ reconstruction (image size $5\times 5$\,nm$^2$ after correction, tunneling voltage $U=0.50$\,V, current $I=200$\,pA). (b) Structure of $k-$space, obtained by fast Fourier transformation, for the image in the panel a. The radius of the circle is equal to $4\pi\sqrt{3}/b =18.89$\,nm$^{-1}$ and it corresponds to the absolute values of the $k$ vectors, describing the two-dimensional periodicity for hexagonal $1\times 1$ lattice; where $b=a/\sqrt{2}=0.384\,$nm is the lattice constant for the Si(111)$1\times 1$ lattice.}
\label{Fig-lateral}
\end{figure*}

The same reconstruction Si(111)$7\times 7$ with pronounced two-dimensional periodicity can be used to find a mean correction factor $k^{\,}_{\|}\simeq 1.14 \pm 0.02$ for scanning tunneling measurements in the lateral direction (figure~\ref{Fig-lateral}). This helps us to convert nominal values $x'$ and $y'$, recorded by the STM Control Unit, into real dimensions
\begin{equation}
x = x'\cdot k^{\,}_{\|} \qquad \mbox{and} \qquad  y = y'\cdot k^{\,}_{\|}.
\label{Eq:lateral-calibration}
\end{equation}

\bigskip

\section*{References}

\begin{enumerate} [label={[S.\arabic*]},itemsep=1 mm]

\item \label{Takayanagi}
Takayanagi, K.;  Tanishiro, Y.;  Takahashi, S.;  Takahashi M.
Structure analysis of Si(111)$-7\times7$ reconstructed surface by transmission electron diffraction. Surface Science. \textbf{1985}, vol.\,\emph{164}, p. 367-392.

\item \label{Tong}
Tong, S. Y.; Huang, H.; Wei, C. M.;  Packard, W. E.; Men, F. K.; Glander, G.; Webb M. B.
Low-energy electron diffraction analysis of the Si(111)7$\times$7 structure. Journal of Vacuum Science and Technology A. \textbf{1988}, vol. \emph{6}, p. 615–624.

\item \label{Oura}
Oura, K.; Lifshits, V. G.;  Saranin, A. A.;  Zotov, A. V.; Katayama M. Surface Science: An Introduction. Springer-Verlag Berlin Heidelberg New York, \textbf{2003}; Springer Science \& Business Media, \textbf{2013}.

\item \label{UserManual}
The LT STM User's Guide (including LT STM-QPlus AFM). Omicron Nanotechnology GmbH. Version 3.4, \textbf{2011}.

\item \label{Vandervoort}
Vandervoort, K. G.;  Zasadzinski, R. K.; Galicia, G. G.; Crabtree, G. W. Full temperature calibration from 4 to 300 K of the voltage response of piezoelectric tube scanner PZT-5A for use in scanning tunneling microscopes. Review of Scientific Instruments. \textbf{1993}, vol.\,\emph{64}, 896–899.

\end{enumerate}

\end{document}